\documentclass[12pt,oneside, reqno]{amsart}
\usepackage{amsaddr}
\usepackage{amsthm}
\usepackage{enumitem}
\usepackage{epsfig}
\usepackage{geometry}
\usepackage{url}
\usepackage{multirow}

\makeatletter

\newcommand{\Rmnum}[1]{\expandafter\@slowromancap\romannumeral #1@}

\def\subsection{\@startsection{subsection}{2}%
\z@{.5\linespacing\@plus.7\linespacing}{.3\linespacing}%
{\normalfont\bfseries\noindent}}
\def\subsubsection{\@startsection{subsubsection}{3}%
\z@{.5\linespacing\@plus.7\linespacing}{.3\linespacing}%
{\normalfont\itshape\noindent}}
\begin{document}

\title{A Coherent Ising Machine Based On Degenerate Optical Parametric Oscillators}
\author{\vspace{-0.1in} \Small Zhe Wang$^{1}$, Alireza Marandi$^{1,2}$, Kai Wen$^1$,}
\author{\vspace{-0.4in} \Small Robert L. Byer$^1$ \and Yoshihisa Yamamoto$^{1,2}$}
\address{\Small 1. E. L. Ginzton Laboratory, Stanford University, Stanford, California 94305, USA\\
2. National Institute of Informatics, Hitotsubashi 2-1-2, Chiyoda-ku, Tokyo 101-8403, Japan}

\begin{abstract}
A degenerate optical parametric oscillator network is proposed to solve the NP-hard problem of finding a ground state of the Ising model. The underlying operating mechanism originates from the bistable output phase of each oscillator and the inherent preference of the network in selecting oscillation modes with the minimum photon decay rate. Computational experiments are performed on all instances reducible to the NP-hard MAX-CUT problems on cubic graphs of order up to 20. The numerical results reasonably suggest the effectiveness of the proposed network.
\end{abstract}

\maketitle

\section{Introduction}

The Ising model is a mathematical abstraction of spin glasses, which are disordered magnetic systems composed of competitively interacting, i.e. frustrated spins. For a spin glass with $N$ spins, the model allows each spin to have two states $\sigma_{j} = \pm 1$, and expresses its energy in terms of the Ising Hamiltonian
\begin{equation}\label{eq:ising_ham}
H = - \sum \limits_{1 \leq j < l \leq N} J_{jl}\sigma_{j}\sigma_{l},
\end{equation}
where $J_{jl}$ denotes the coupling coefficient between the $j$-th and the $l$-th spin. The coupling is called ferromagnetic if $J_{jl} > 0$ and antiferromagnetic if $J_{jl} < 0$.

Unlike ordered systems in ideal crystals, spin glasses possess peculiar properties around critical temperatures such as spin freezing, existence of a cusp in magnetic susceptibility, remanence and hysteresis \cite{BY86}. The Ising model serves as a useful prototype for studying these unusual behaviors \cite{D00}. Many of these properties are considered to be arising from the absence of global magnetic order of the low-lying energy states in the system \cite{NS03}. Therefore, an Ising machine capable of efficiently outputting the ground state spin configurations of the Ising Hamiltonian is highly demanded. 

The presence of an Ising machine is also desirable to other diverse areas including computer science \cite{MP87}, biology \cite{S92} and information processing \cite{N01}. In fact, the problem of finding a ground state of an Ising Hamiltonian, to be referred to as the \textit{Ising problem} in the following, is in the NP-hard category in computational complexity \cite{B82}. Many combinatorial optimization problems arising in various areas belong to the same category. For instance, the MAX-CUT problem on an undirected graph $G = (V,E)$ with an edge weight function $w : E \rightarrow \mathbb{R}$ is one of them, where $V$ and $E$ denote the sets of vertices and edges respectively. The goal of the MAX-CUT problem is to find a cut $(S,V\setminus S)$ such that the sum of the weight of the edges with one endpoint in $S$ and the other in $V \setminus S$ is maximized over all possible cuts \cite{K72}. Let $w_{jl} = w_{lj}$ be the edge weight if $(j,l) \in E$ and $w_{jl} = 0$ if $(j,l) \notin E$, and $\sigma_{j} = +1$ if the $j$-th vertex is in $S$ and $\sigma_{j} = -1$ if not. The weight of a cut $S$ is thus given by
\[ 
	w(S) = \sum\limits_{j \in S, l \in V\setminus S}w_{jl} = \frac{1}{4}\sum\limits_{j,l \in V} w_{jl} - \frac{1}{4}\sum\limits_{j,l \in V} w_{jl}\sigma_j\sigma_l.
\]
When the coupling coefficient $J_{jl}$ in the Ising model is chosen to be $-w_{jl}$, any maximum cut of the given graph can be converted to a ground state of the corresponding Ising problem and vice versa. Likewise, NP-hard problems are reducible to each other by polynomial transformations. 

Once a polynomial time algorithm is available for one member of the NP-hard problems, all the problems in this category can be solved efficiently \cite{GJ79}. So far no such method is known and these problems are commonly believed to be intractable. Nevertheless, many attempts have still been undertaken to tackle them. The simulated annealing algorithm is designed by mimicking the thermal annealing procedure in metallurgy \cite{KG83}. Making use of the quantum tunneling process, quantum annealing technique was also formulated \cite{KN98} and is shown to have superior performance over simulated annealing \cite{SM02}. As a variant of quantum annealing, quantum adiabatic computation was devised according to the adiabatic theorem of quantum mechanics \cite{FG01}, with computational power equivalent to that of a quantum computer based on unitary gates \cite{DM01,AV07}. Despite the fact that none of these methods are generally proven to be efficient, taking advantage of fundamental principles in physics has shed new light on solving NP-hard problems. In this regard, it is worthwhile studying the computational ability of other promising physical systems to search for alternative approaches.

Lasers are open dissipative systems that undergo second-order phase transition at the oscillation threshold. Potentially oscillating modes in a laser compete for the available gain and reduce the gain accessible to other modes due to the cross-saturation effect \cite{S86}. Since the mode with the minimum threshold gain is more likely to be excited first, it has the edge over other modes to spontaneously emerge through the mode competition. This phenomenon is to be referred to as the \textit{minimum gain principle} in the following. It is demonstrated that the overall photon decay rate in a mutually injection-locked laser network can be engineered to be in the form of an Ising Hamiltonian \cite{UT11}. Each laser in the network is polarization degenerate, and it represents Ising spin $+1$ in the case that right circularly polarized photons outnumber left circularly polarized photons and Ising spin $-1$ in the opposite case. Each combination of the polarization of all lasers is considered to be a global mode of the whole network. Numerical simulations have shown evidence in favor of the minimum gain principle. For the select Ising problems, the network is likely to oscillate in global modes with the minimum photon decay rate. Moreover, the transient time of the network to the steady state is estimated to be determined by the mutual injection signals among the lasers, which does not scale with the number of spins in the Ising problem \cite{TU12}.  

Degenerate parametric oscillators are also open dissipative systems that experience second-order phase transition at the oscillation threshold \cite{WL71}. Due to the phase-sensitive amplification, however, an oscillator operating above the threshold can only oscillate with one of two possible phases. In the early development of digital computers, logic circuits were built from electrical oscillators of this type by using the bistability of their output phases \cite{G59}. The phase that an oscillator would take from the two equally preferred outcomes is randomly determined by the noise. In the case of a degenerate optical parametric oscillator (OPO), quantum noise associated with the optical parametric down conversion during the oscillation build-up takes the charge \cite{LY61}. Based on this property, a quantum random number generator was implemented by taking XOR of the phases of two independently oscillating degenerate OPOs \cite{ML12}. In these applications, some computational abilities of parametric oscillators were explored, but the oscillators were merely treated as individual binary digits.

This paper is concerned with collective behaviors of a degenerate OPO network. It is the first proposal to build a coherent Ising machine from phase-sensitive oscillators like degenerate OPOs as opposed to phase-insensitive oscillators like lasers. Moreover, the proposed Ising machine achieves substantial improvement in performance compared to the aforementioned one based on lasers, as will be mentioned in section \ref{sec:6}. 

In the proposed network, each degenerate OPO is identified as an Ising spin by its binary output phase. Each particular overall phase configuration of the network becomes a global mode and represents an Ising spin state. In order to solve an Ising problem, the output fields of the degenerate OPOs are coherently injected to others with the amplitudes and phases governed by the coupling coefficients in the given problem. Under appropriate implementation, the overall photon decay rate of the global mode is proportional to the energy of the corresponding Ising spin state. Since the minimum gain principle is also applicable to the mode selection, the network will probably give a solution to the NP-hard Ising problem.

The paper is structured as follows. Section \ref{sec:2} prepares the theoretical groundwork for examination of the proposed network through the study of a single degenerate OPO. Section \ref{sec:3} presents the dynamical equations of the network, and analyzes its steady state properties which are essential for the computational ability as an Ising machine. To illustrate the situations that the network solves the Ising problem, the case of two coupled degenerate OPOs are discussed in section \ref{sec:4}. Performance of the network is evaluated by conducting computational experiments against the NP-hard MAX-CUT problem on cubic graphs \cite{Y78} in section \ref{sec:5}. Finally, section \ref{sec:6} concludes the paper.

\section{A Single Degenerate OPO}\label{sec:2}

A degenerate OPO consists of a second order nonlinear crystal placed in an optical cavity. Under the drive of a coherent external pump $F_p$ at frequency $\omega_p$, a pump field is excited inside the cavity. Due to the second order susceptibility of the nonlinear crystal, a signal field at frequency $\omega_s$ is generated from the pump field such that $\omega_{p}=2\omega_{s}$. Assume that $F_p$ is classical and its phase is the reference phase of the oscillator. The Hamiltonian of a degenerate OPO is hence given by
\begin{equation}\label{eq:opo_ham}
\begin{array}{l}
	\displaystyle H = H_{0} + H_{\mathrm{int}} + H_{\mathrm{irr}} \\
	\displaystyle H_{0} = \hbar\omega_{s}\hat{a}_s^\dagger\hat{a}_s + \hbar\omega_{p}\hat{a}_p^\dagger\hat{a}_p \\
	\displaystyle  H_{\mathrm{int}} = i\hbar\frac{\kappa}{2}\left(\hat{a}_s^{\dagger 2}\hat{a}_p  - \hat{a}_s^2\hat{a}_p^\dagger \right) \\
	\displaystyle \hspace{0.5in} + i\hbar\sqrt{\gamma_{p}}\left(\hat{a}_p^\dagger F_{p}\mathrm{e}^{-i\omega_{p}t} - \hat{a}_pF_{p}\mathrm{e}^{i\omega_{p}t} \right)\\
	\displaystyle H_{\mathrm{irr}} = i\hbar\sqrt{\gamma_{s}}\left(\hat{a}_s^\dagger\hat{B}_s - \hat{a}_s\hat{B}_s^\dagger\right) \\
	\displaystyle \hspace{0.5in} + i\hbar\sqrt{\gamma_{p}}\left(\hat{a}_p^\dagger\hat{B}_p - \hat{a}_p\hat{B}_p^\dagger\right).
\end{array}
\end{equation}
Here, $H_0$ represents the energies of the signal and the pump fields inside the cavity, where $\hat{a}_s^\dagger, \hat{a}_s$ are the creation and annihilation operators for the signal field,  and $\hat{a}_p^\dagger, \hat{a}_p$ are the counterparts for the pump field. Also, the first term in $H_{\mathrm{int}}$ describes the nonlinear coupling between the signal and the pump fields, where $\kappa$ is the parametric gain due to the second order susceptibility of the nonlinear crystal; whereas the second term shows the excitation of the internal pump field by the external pump. Finally, $H_{\mathrm{irr}}$ denotes the irreversible interaction between cavity fields and the reservoir, where $\hat{B}_s, \hat{B}_p$ are reservoir operators with continuous spectra in the frequency domain, and $\gamma_s, \gamma_p$ are the signal and the pump photon decay rates from the cavity \cite{DM80}.

From the Hamiltonian in eq.(\ref{eq:opo_ham}), the Heisenberg-Langevin equations of a degenerate OPO can be derived as
\[ \begin{array}{l}
	\displaystyle \frac{d}{d\tau}\hat{A}_{s}  =  -\frac{\gamma_s}{2}\hat{A}_s + \kappa\hat{A}_s^\dagger\hat{A}_p + \sqrt{\gamma_s}\hat{f}_s \smallskip \\
	\displaystyle \frac{d}{d\tau}\hat{A}_{p}  =  -\frac{\gamma_p}{2}\hat{A}_p - \frac{\kappa}{2}\hat{A}_s^2 + \sqrt{\gamma_p}\left(F_p + \hat{f}_p\right),
\end{array}\]
where $\hat{A}_s = \hat{a}_s\mathrm{e}^{i\omega_s\tau}$, $\hat{A}_p = \hat{a}_p\mathrm{e}^{i\omega_p\tau}$ denote the slowly varying signal and pump operators in the rotating frame, and $\hat{f}_s, \hat{f}_p$ are the time-dependent noise operators to the signal and the pump fields, respectively \cite{O00}.

The c-number Langevin equations subsequently follow from converting each operator in the above equations to a complex number. Moreover, under the condition $\gamma_s \ll \gamma_p$, the pump field can be adiabatically eliminated since it immediately follows the change of the signal field. With this slaving principle, the c-number Langevin equations reduce to a single stochastic differential equation     
\begin{equation}\label{eq:nde_sgl_opo}
\begin{array}{l}
	\displaystyle \frac{d}{d\tau}\tilde{A_{s}} = -\frac{\gamma_s}{2}\tilde{A_s} + \kappa (\frac{2}{\sqrt{\gamma_p}}F_{p} - \frac{\kappa}{\gamma_p}\tilde{A_{s}}^{2})\tilde{A_{s}}^{*} + \frac{2\kappa}{\sqrt{\gamma_p}}\tilde{A_{s}}^{*}\tilde{f_p} + \sqrt{\gamma_s}\tilde{f_s},
\end{array}
\end{equation}
which describes the dynamics of the complex amplitude $\tilde{A_s}$ of the signal field. Here, the superscript $^{*}$ denotes the operation of complex conjugate, and $\tilde{f}_s = f_{s,1} + if_{s,2}$, $\tilde{f}_p = f_{p,1} + if_{p,2}$ are quantum noises to the signal and pump fields respectively, whose real and imaginary components $f_{k,j}, \; k \in \{s,p\}, \; j \in \{1,2\}$ are independent white Gaussian noises with the ensemble averaged means and correlations satisfying
\[ \begin{array}{l}
	\displaystyle \langle f_{k,j}(\tau)\rangle = 0, \\
	\displaystyle \langle f_{k,j}(\tau)f_{k',j'}(\tau ')\rangle = \frac{1}{4}\delta_{kk'}\delta_{jj'}\delta(\tau-\tau'). 
\end{array}\]
The real and imaginary parts of the complex amplitude are called the in-phase and quadrature components of the signal field, respectively. Dynamical equations for the in-phase and the quadrature components can be easily derived from eq.(\ref{eq:nde_sgl_opo}).

\begin{figure}[t]
\centering
\epsfig{figure = 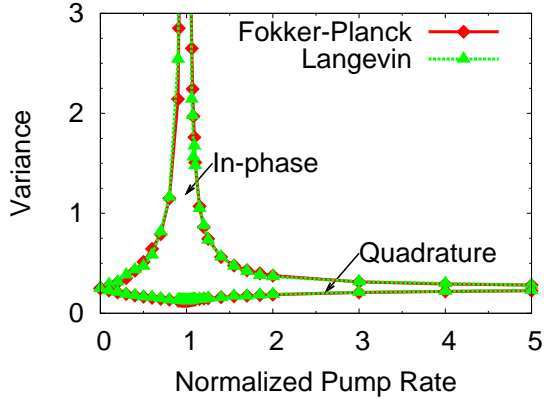, width = 2in, angle = -90} 
\caption{Variances of the in-phase and the quadrature components of the signal field at the steady state at different pump levels when $\gamma_s = 1$, $\gamma_p = 100$, and $\kappa = 0.1$. The numerical values are calculated from two theoretical models: the quantum mechanical Fokker-Planck approach and the classical c-number Langevin approach.}
\label{fig:var_sgl_dopo}
\end{figure}

Despite the classical nature of the c-number Langevin equation, it produces identical prediction on statistical quantities of the signal field as a quantum mechanical approach does. Fig.\ref{fig:var_sgl_dopo} displays the second central moments of the in-phase and the quadrature components at the steady state. The normalized pump rate is defined as $p = F_p/F_{th}$, where $F_{th} = {\gamma_s\sqrt{\gamma_p}}/{4\kappa}$ is the threshold external pump flux. It can be easily seen that the c-number Langevin approach successfully predicts the generation of squeezed states around the threshold \cite{MW81}. Meanwhile, the numerical values are in good agreement with the results calculated from the quantum mechanical Fokker-Planck equation obtained from the generalized $P$-representation \cite{DM80}. Therefore, this classical approach is adopted for the investigation of the proposed degenerate OPO network.

\section{A Degenerate OPO Network}\label{sec:3}

In order to solve an Ising problem with $N$ spins, a network composed of $N$ degenerate OPOs needs to be constructed. Each degenerate OPO in the network corresponds to an Ising spin, and its signal output is coherently injected to another according to the coupling coefficient between the spins involved. Let $\xi_{jl}/2$ denote the scaling factor for the complex signal field $\tilde{A_j}$ of the $j$-th degenerate OPO when it is coupled to the $l$-th degenerate OPO. Since the coupling coefficients in the Ising problem are always symmetric, it follows that $\xi_{jl} = \xi_{lj}$. By further adding terms representing the mutual coupling of the signal fields to eq.(\ref{eq:nde_sgl_opo}), the c-number Langevin equations of the network are obtained as
\begin{equation}\label{eq:nde_cplx}
\begin{array}{l}
	\displaystyle \frac{d}{d\tau}\tilde{A_{j}} = -\frac{\gamma_s}{2}\tilde{A_j} + \kappa (\frac{2}{\sqrt{\gamma_p}}F_{p} - \frac{\kappa}{\gamma_p}\tilde{A_{j}}^{2})\tilde{A_{j}}^{*} + \frac{1}{2}\sum\limits_{l = 1, l \neq j}^{N} \xi_{jl}\gamma_{s}\tilde{A_{l}} \\
	\displaystyle \hspace{2.7in} + \frac{2\kappa}{\sqrt{\gamma_p}}\tilde{A_{j}}^{*}\tilde{f}_{p,j} + \sqrt{\gamma_s}\tilde{f}_{s,j}
\end{array}
\end{equation}
where $\tilde{f}_{s,j}$ and $\tilde{f}_{p,j}$ are the associated quantum noises to the $j$-th degenerate OPO.

For a single degenerate OPO pumped above the threshold, eq.(\ref{eq:nde_sgl_opo}) implies that the mean of its quadrature component at the steady state is 0. Therefore, the phase of the oscillating field on average is either 0 or $\pi$ determined by the sign of its in-phase component. It is extremely favorable if all degenerate OPOs in the coupled network operating above the threshold still possess the bistability of their output phases. This is because the phase configuration of the network can be naturally converted to an Ising state by assigning $\sigma_{j} = +1$ to the $j$-th degenerate OPO if its in-phase component is positive, or $\sigma_{j} = -1$ if negative. Due to the mutual coupling, however, it is not self-evident that this feature is available to the network. In the following, required conditions are explored through analyzing steady state properties of the network.

Dynamical equations for the in-phase components $C_{j}$ and the quadrature components $S_{j}$ of the complex signal amplitudes $\tilde{A_{j}}$ provide an equivalent description of the network as eq.(\ref{eq:nde_cplx}). Since the theoretical investigation is mainly interested in the mean signal fields at the steady state, noise terms are neglected in the dynamical equations. For the ease of analysis, normalized equations for the in-phase and quadrature components
\begin{equation}\label{eq:nde_inp_qua_norm}
\begin{array}{l}
\displaystyle \frac{d}{dt}c_{j}  =  (-1 + p - (c_{j}^{2}+s_{j}^2))c_{j} + \sum\limits_{l = 1, l \neq j}^{N} \xi_{jl}c_{l} \\
\displaystyle \frac{d}{dt}s_{j}  =  (-1 - p - (c_{j}^{2}+s_{j}^2))s_{j} + \sum\limits_{l = 1, l \neq j}^{N} \xi_{jl}s_{l}
\end{array}
\end{equation}
are utilized, where $t = \gamma_s\tau/2$ is the unitless time normalized to twice of the signal photon cavity lifetime, and $c_{j} = C_{j}/A_{s}$, $s_{j} = S_{j}/A_{s}$ are normalized in-phase and quadrature components where $A_{s} = \sqrt{\gamma_s\gamma_p/2\kappa^2}$ is the signal amplitude of a single degenerate OPO when $p = 2$.  The above equations indicate that the dynamics of the network are influenced by the values of $p$ and $\xi_{jl}$. 

\subsection{Oscillation Threshold} 

The in-phase and the quadrature components of the $N$ coupled degenerate OPOs satisfy
\begin{equation} \label{eq:inph_quad_ss}
\begin{array}{l}
\displaystyle c_{j}^{3} + (1 - p + s_{j}^{2})c_{j} - \sum\limits_{l = 1, l \neq j}^{N} \xi_{jl}c_{l} = 0 \\
\displaystyle s_{j}^{3} + (1 + p + c_{j}^{2})s_{j} - \sum\limits_{l = 1, l \neq j}^{N} \xi_{jl}s_{l} = 0
\end{array}
\end{equation}
at the steady state. The oscillation threshold of the network is defined as the normalized pump rate $p_{th}$ above which the network cannot arrive at the trivial steady state $c_j = s_j = 0$, $\forall j \in \{1,2,\dots,N\}$. For any hermitian matrix $A$, let $\lambda_{\mathrm{min}}(A)$ and $\lambda_{\mathrm{max}}(A)$ be the smallest and the largest eigenvalues, respectively. Since the largest eigenvalue of the corresponding Jacobian matrix
\[ \displaystyle J_0 =  \left( \begin{array}{cc}
-(1-p)I_N - G & 0 \\
0 & -(1+p)I_N - G
\end{array} \right) \]
has to be positive above the threshold, the threshold of the network is obtained as
\begin{equation}\label{eq:thres_cond}
	p_{th} = 1 + \lambda_{\mathrm{min}}(G) < 1,
\end{equation}
where $I_N$ is the $N \times N$ identity matrix and
\[ \displaystyle G = \left( \begin{array}{cccc}
0 & -\xi_{12} & \dots & -\xi_{1N}\\
-\xi_{21} & 0 & \dots & -\xi_{2N}\\
\vdots & \vdots & \ddots & \vdots \\
-\xi_{N1} & -\xi_{N2} &\dots & 0 
\end{array} \right) \]  
is the hermitian matrix showing the coupling relation of the network. Since $\mathrm{Tr}(G) = 0$, it follows that $\lambda_{\mathrm{min}}(G) < 0$ and the threshold of the network is lower than that of an individual degenerate OPO. This phenomenon is similar to the so-called self-ignition effect well known in the study of neural networks \cite{KA92}.
   
\subsection{Quadrature Components}

It follows from the quadrature component equations in eq.(\ref{eq:inph_quad_ss}) that
\[ \sum\limits_{j = 1}^{N}s_{j}^{4} + \sum\limits_{j = 1}^{N}a_{j}s_{j}^{2} - \sum\limits_{j = 1}^{N}\sum\limits_{l = 1, l \neq j}^{N} \xi_{jl}s_{j}s_{l} = 0, \]
where $a_{j} = 1 + p + c_{j}^{2}, j \in \{1,2, \dots, N\}$. The last two terms of the above equation is in the quadratic form of the matrix
\[ \displaystyle Q = \left( \begin{array}{cccc}
a_{1} & -\xi_{12} & \dots & -\xi_{1N}\\
-\xi_{21} & a_{2} & \dots & -\xi_{2N}\\
\vdots & \vdots & \ddots & \vdots \\
-\xi_{N1} & -\xi_{N2} &\dots & a_{N} 
\end{array} \right). \]
If $Q$ is positive-definite, $s_{j} = 0$, $\forall j \in \{1,2, \dots, N\}$ will be the only possible solution to eq.(\ref{eq:inph_quad_ss}) for the quadrature components. Since the dynamics of the in-phase components are affected by the square of the quadrature components, the local behaviors of the in-phase and the quadrature components around this solution can be separated. The corresponding Jacobian matrix for the quadrature components is $J_{s} = - Q$. Thus, a steady state with all quadrature components being 0 is stable if its corresponding Jacobian matrix for the in-phase components is negative-definite.

The smallest eigenvalue of $Q$ is still unknown without solving eq.(\ref{eq:inph_quad_ss}) to obtain the steady state values of the in-phase components. However, a lower bound can be easily evaluated. The hermitian matrix $Q$ can be written as the sum of the hermitian matrix $G$ and a diagonal matrix whose diagonal components are $a_1, a_2, \dots, a_N$. From Weyl's theorem \cite{F93}, the eigenvalues $\lambda(Q)$ are bounded as
\[\lambda_{\mathrm{min}}(G) + \min\limits_{j}a_{j} \leq \lambda(Q) \leq \max\limits_{j}a_{j} + \lambda_{\mathrm{max}}(G). \]
Since $\lambda_{\mathrm{min}}(G) < 0$, a sufficient condition for $Q$ to be positive-definite is
\begin{equation} \label{eq:quad_0}
|\lambda_{\mathrm{min}}(G)| < 1 + p \leq \min\limits_{j}a_{j}.
\end{equation}
The above requirement together with the oscillation threshold in eq.(\ref{eq:thres_cond}) establish guidelines for choosing appropriate combinations of the normalized pump rate $p$ and the coupling strength $\xi_{jl}$ to identify the degenerate OPOs in the network as Ising spins. Yet it is noteworthy that combinations of $p$ and $\xi_{jl}$ not satisfying eq.(\ref{eq:quad_0}) may also be preferred in some cases because of its sufficient nature.

\subsection{Overall Photon Decay Rate}

The term $p-\left(c_j^2+s_j^2\right)$ in eq.(\ref{eq:nde_inp_qua_norm}) represents the saturated gain for the $j$-th degenerate OPO. At the steady state, the total saturated gain of the network equals to the overall photon decay rate $\Gamma$. In the case that all the quadrature components of the degenerate OPOs are 0, $\Gamma = \sum_{j = 1}^{N}\left(p-c_j^2\right)$.

When mutual coupling of the degenerate OPOs is weak enough and $p > 1$, the in-phase component $c_j$ in eq.(\ref{eq:inph_quad_ss}) can be expressed in the formal expansion
\[
	c_j = c_j^{(0)} + \epsilon c_j^{(1)} +\epsilon^2 c_j^{(2)} + \dots, \]
where $\epsilon = \max_{1 \leq j,l \leq N}|\xi_{jl}|$, according to the perturbation theory. Each term $c_j^{(n)}, n \geq 0$ can be analytically obtained by substituting the above expansion to eq.(\ref{eq:inph_quad_ss}) and setting the coefficient of the $\epsilon^n$ term to be 0. The 0-th order term $c_j^{(0)} = \pm \sqrt{p-1}$ is the signal amplitude of the $j$-th degenerate OPO operating above the threshold when there is no mutual coupling. Since the formal expansion for $c_j$ can be viewed as a local modification to $c_j^{(0)}$, the Ising spin value $\sigma_j$ that the $j$-th degenerate OPO represents equals to $\mathrm{sgn}(c_j^{(0)})$. The overall photon decay rate is thus given by
\begin{equation}\label{eq:tot_gamma_Ising}
	\Gamma = N - \sum\limits_{1 \leq j \neq l \leq N} \xi_{jl}\sigma_j\sigma_l + O\left(\frac{\epsilon^3N^4}{(p-1)^3}\right),
\end{equation}
where the higher order correction term is evaluated in the case when coupling of the same strength exists between any two of the degenerate OPOs. For a particular phase configuration of the degenerate OPOs, the difference of the overall photon decay rates between cases with and without mutual coupling is exactly the energy of its corresponding spin configuration in an Ising problem where the coupling coefficients between spins are $2\xi_{jl}$. Therefore, a global mode that achieves the minimum $\Gamma$ provides a ground state to the Ising problem.

Given any Ising problem, scaling all the coupling coefficients by the same positive factor does not change its solutions. In this regard, some degree of flexibility is available in choosing the coupling strength of the network. For a fixed pump rate, the gaps among the overall photon decay rates of different modes decrease with weaker coupling strength. As a consequence, the possibility that the network evolves into steady states corresponding to excited states of the Ising Hamiltonian may be increased. On the other hand, the mapping from the overall photon decay rate to the Ising Hamiltonian becomes more inaccurate when the coupling strength gets stronger, which may also cause probable errors in solving the Ising problem. This intuitive observation indicates that the choice of $p$ and $\xi_{jl}$ can significantly influence the performance of the network as an Ising machine. As an example, the system of two coupled degenerate OPOs is examined in the next section. 

\section{Two Coupled Degenerate OPOs}\label{sec:4}

Dynamical behaviors of two coupled degenerate OPOs offer meaningful insight into how the values of the normalized pump rate $p$ and the coupling coefficient $\xi$ can change the candidate solutions that the system finds to an Ising problem. Due to its relative simplicity, many properties of the system can be studied analytically. The threshold of the system is $p_{th} = 1 - \left|\xi\right|$ by eq.(\ref{eq:thres_cond}). When $\left|\xi\right| \leq 1+p$, all the quadrature components of the system are 0 at the steady state according to the sufficient condition in eq.(\ref{eq:quad_0}). Expressions of the in-phase components $c_1$ and $c_2$ at the steady state can also be easily obtained, which further allows the investigation of their linear stability. The results are summarized in Table \ref{tbl:ss_two_dopo}.

\begin{table}[t]
\caption{Linear stability of steady states with all quadrature components being 0 of two coupled degenerate OPOs. Here, $c_{g}^{2} = p-1+\xi$, $c_{u}^{2} = p-1-\xi$, $c_{s}^2 = (p-1\pm\sqrt{(p-1)^{2}-4\xi^{2}})/2$, $c_{a}c_{s} = -\xi$, and NA means the corresponding steady state does not exist.}
\centering
\renewcommand{\arraystretch}{1.1}
\begin{tabular}{c|c|cccc}
\hline\hline
 & & \multicolumn{4}{|c}{$(c_{1},c_{2})$} \\
$p$ & $\xi$ & $(0,0)$ & $(c_{g},c_{g})$ & $(c_{u},-c_{u})$ & $(c_{s},c_{a})$ \\
\hline
\multirow{3}{*}{$p \leq 1$} & $(1-p,1+p)$ & unstable & stable & NA & NA\\
 & $(p-1,1-p)$ & stable & NA & NA & NA \\
 & $(-1-p,p-1)$ & unstable & NA & stable & NA \\
\cline{1-6}
\multirow{5}{*}{$p > 1$} & $(p-1,p+1)$ & unstable & stable & NA & NA\\
 & $(\frac{p-1}{2},p-1)$ & unstable & stable & unstable & NA\\
 & $(-\frac{p-1}{2},\frac{p-1}{2})$ & unstable & stable & stable & unstable \\
 & $(1-p,-\frac{p-1}{2})$ & unstable & unstable & stable & NA\\
 & $(-1-p,1-p)$ & unstable & NA & stable & NA\\     
\hline
\end{tabular}
\label{tbl:ss_two_dopo}
\end{table}

\begin{figure}[t]
\centering
\epsfig{figure = 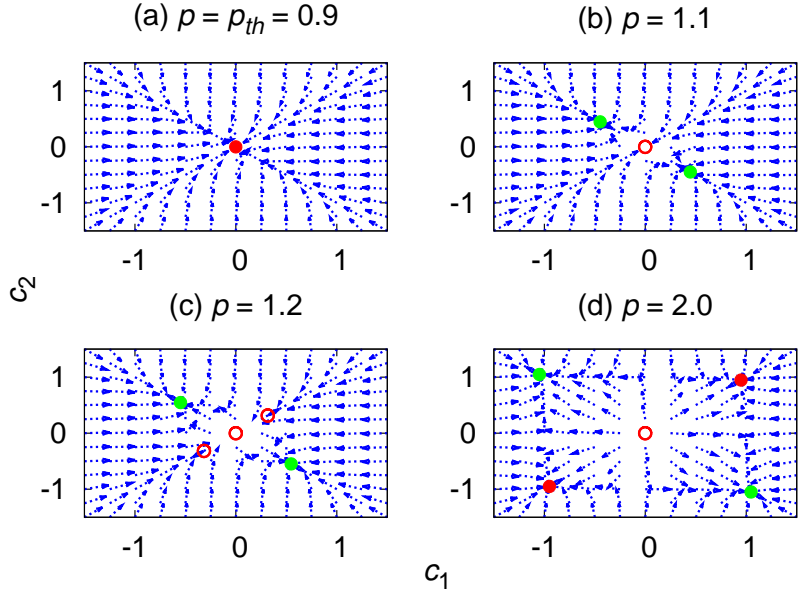, width = 3in, angle = -90} 
\caption{Phase diagrams for the in-phase components of two coupled degenerate OPOs when $\xi=-0.1$. The dots and circles mean stable and unstable steady states, while the color green and red denote correct and incorrect solutions, respectively.}
\label{fig:phase_diagram}
\end{figure}

As shown in Table \ref{tbl:ss_two_dopo}, possible steady states that the system can evolve into depends considerably on the values of $p$ and $\xi$. Indeed, it is the phase diagram of the in-phase components being modified. Fig.\ref{fig:phase_diagram} displays the variation of the phase diagram with $p$ when $\xi$ is fixed:
\begin{enumerate}[label=(\alph*)]
	\item When the system is pumped below or at the threshold, the only possible steady state is the trivial one $c_1 = c_2 = 0$.
	\item As the pump rate is slightly increased, the trivial steady state becomes unstable while additional two stable steady states are developed. For the case being displayed, the two degenerate OPOs have the same amplitude but opposite phases at both the two newly developed stable steady states. Therefore, they correspond to the correct solutions to the Ising problem of two spins with an antiferromagnetic coupling. Since these two steady states are the only stable ones available, the system can always arrive at the correct answers irrespective of its initial state.
	\item As the pump rate is further increased, two more steady states begin to exist in the system. Since the two degenerate OPOs have the same amplitude and phase at these two steady states, they correspond to the wrong solutions to the Ising problem. However, at this pump level these two steady states are unstable ones so that the system can still solve the Ising problem with certainty.
	\item When the pump rate is raised up over the critical point $1 + 2|\xi|$ according to Table \ref{tbl:ss_two_dopo}, the steady states corresponding to the wrong answers also become stable. Since the system can evolve into these steady states as well, errors may occur for the system in solving the Ising problem.
\end{enumerate}

\begin{figure}[t]
\centering
\epsfig{figure = 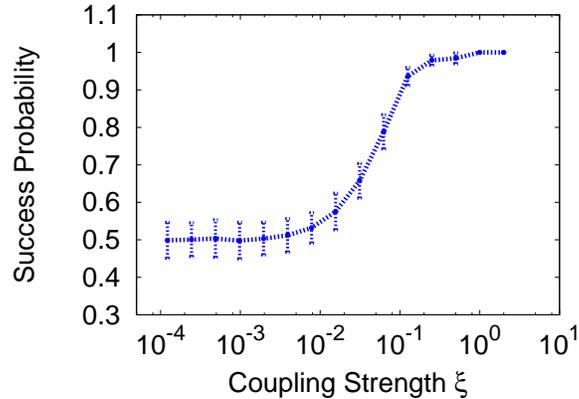, width = 2in, angle = -90} 
\caption{The success probability of two coupled degenerate OPOs in solving the Ising problem of two spins with a ferromagnetic coupling when $p = 2.0$. The error bars are standard deviations.}
\label{fig:cor_vs_xi}
\end{figure}

As a result, the choice of $p$ and $\xi$ affects to a large extent the efficiency of the system as an Ising machine. Fig.\ref{fig:cor_vs_xi} displays the $\xi$ dependence of its success probability when $p$ is fixed in solving the Ising problem of two spins with a ferromagnetic coupling. A candidate answer from the system can be obtained by numerically solving its dynamical equations under a random noise input. The details of the numerical implementation are described in the next section. For each coupling coefficient, 100 groups of 100 random trials are conducted to estimate the success probability and the standard deviation. The success probability asymptotically approaches 0.5 as the coupling strength decreases. This is because in the limit of no coupling each degenerate OPO can choose its phase independently so that correct and incorrect solutions are equally likely. On the other hand, when $\xi$ is above the critical point $(p-1)/2$, the system outputs a correct solution for every trial.

\section{Computational Experiments}\label{sec:5}

Performance of the proposed degenerate OPO network as an Ising machine is tested against the NP-hard MAX-CUT problem on cubic graphs. All the cubic graphs of order up to 20 are investigated. The reason for using these small instances is because the correctness of the output from the network can be verified by checking all possible cuts by brute force. The factor that limits the maximum input size for the current investigation is the total number of instances of each graph order, which is irrelevant to the properties of the network. As shown in Table \ref{tbl:cubic_graph}, it grows even faster than an exponential function with respect to the input size \cite{R96}. Thus, the time it takes to exhaust all cubic graphs of the same order climbs commensurately.

\begin{table}[t]
\renewcommand{\arraystretch}{1.2}
\caption{Number of cubic graphs.}
\centering
\begin{tabular}{cccccccccccc}
\hline\hline 
Order & 4 & 6 & 8 & 10 & 12 & 14 & 16 & 18 & 20 & 22 & 24 \\
\hline
Cubic Graphs & 1 & 2 & 5 & 19 & 85 & 509 & 4060 & 41301 & 510489 & 7319447 & 117940535\\
\hline\hline
\end{tabular}
\label{tbl:cubic_graph}
\end{table}

\subsection{Implementation}

\begin{figure}[t]
\centering
\epsfig{figure = 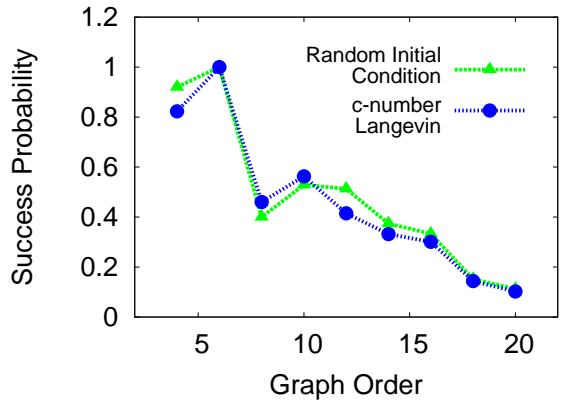, width = 2in, angle = -90} 
\caption{The success probability of the network in solving the MAX-CUT problem on the worst-case instances listed in Table \ref{tbl:worst_case} when $p = 1.1$ and $\xi = -0.1$. The parameters used in the calculation are $\gamma_s = 1$, $\gamma_p = 100$ and $\kappa = 10^{-4}$ for the c-number Langevin approach, and $A_{ini} = 10^{-5}$ for the random initial condition approach.}
\label{fig:suc_pr_vs_node_comp}
\end{figure}

Possible solution outputs from the network can be obtained by solving the c-number Langevin equations in eq.(\ref{eq:nde_cplx}) with the signal field of each degenerate OPO starting from the vacuum state. However, the existence of quantum noise inputs in these equations makes the computational cost of this method relatively expensive. To avoid this issue, an alternative method presented in the following is used instead, which ignores the quantum noise terms and replaces them with a random initial condition. In this way, the differential equations to be dealt with switch over from stochastic ones to deterministic ones, which considerably improves the efficiency of the numerical simulation. More importantly, as shown in Fig.\ref{fig:suc_pr_vs_node_comp}, these two methods provide approximately identical results in terms of the success probability of the network in finding a correct answer when the random initial amplitude of the degenerate OPOs is selected to be the same order of magnitude as the quantum noise strength of the network.

For a cubic graph with $N$ vertices, the $2N$ classical dynamical equations of the in-phase and quadrature components in eq.(\ref{eq:nde_inp_qua_norm}) are solved to obtain possible cuts. The Dormand-Prince method is chosen as the algorithm for the differential equation solver, which allows adaptive integration steplength by evaluating the local truncation error \cite{L91}. In order to simulate the quantum noise, initial conditions to eq.(\ref{eq:nde_inp_qua_norm}) are randomly generated in the neighborhood of the trivial steady state $c_j = s_j = 0$, $\forall j \in \{1,2,\dots,N\}$. In the current simulation, the degenerate OPOs initially have the same normalized amplitude $A_{ini} = 10^{-5}$ but different random phases. The simulation continues until the network approaches a stable steady state. Since the number of equations required to be solved only increases linearly with the number of vertices, this method is also suitable for the investigation of large input size instances.

\subsection{Results}

As a first attempt to solve the MAX-CUT problem using the degenerate OPO network, the normalized pump rate and the coupling coefficient are fixed at $p = 1.1$, $\xi = -0.1$ for all the instances. This choice of $p$ and $\xi$ reflects the realistically achievable experimental condition. In addition, it guarantees that the network is operating above its threshold defined by eq.(\ref{eq:thres_cond}) for all instances. Since the eigenvalues of the hermitian matrix $G$ in eq.(\ref{eq:thres_cond}) for all the cubic graphs are bounded between $-3|\xi|$ and $3|\xi|$, the phases of the degenerate OPOs are either 0 or $\pi$ at the steady state according to the sufficient condition in eq.(\ref{eq:quad_0}).

\begin{figure}[t]
\centering
\epsfig{figure = 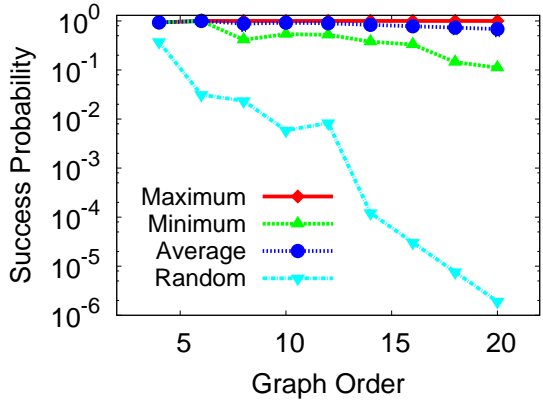, width = 2in, angle = -90} 
\caption{The success probability of the network in solving the MAX-CUT problem on cubic graphs when $p = 1.1$ and $\xi = -0.1$. The maximum, minimum and average success probabilities are evaluated over instances of the same order. The success probability of a random guess for the worst-case instances is also included for comparison. The error bars denote the standard deviations.}
\label{fig:suc_pr_vs_node_p_1.1_xi_0.1}
\end{figure}

The success probability of the network in finding a maximum cut is estimated by examining its approached steady states under 100 random initial conditions. For instances where the success probability is below 0.25 or in the last 10 lowest among the cubic graphs of the same order, additional 10000 trials are conducted to refine the estimation. As shown in Fig.\ref{fig:suc_pr_vs_node_p_1.1_xi_0.1}, the network is able to output a maximum cut with a high success rate for most of the instances. The average success probability for cubic graphs of order 20 is about 0.682. Even in the worst cases where the minimum success probability is attained, the network still substantially outperforms a random guess. For the worst-case instance with 20 vertices listed in Table \ref{tbl:worst_case}, the network amplifies the success rate of picking one of the only 2 correct answers out of $2^{20} \approx 10^6$ candidates by about 60000 times.

\begin{table}[t]
\renewcommand{\arraystretch}{1.4}
\caption{Worst-case success probability $q$ of the network in solving the MAX-CUT problem on cubic graphs when $p = 1.1$ and $\xi = -0.1$. The number of maximum cuts $N_0$ and the number of second largest cuts $N_1$ for the worst-case instances are also included. Here, for a given graph $G = (V,E)$, a subset $S \subseteq V$ and its complement $ V \setminus S$ are considered to be two different cuts.}
\centering
\begin{tabular}{cccccccccc}
\hline\hline 
Order & 4 & 6 & 8 & 10 & 12 & 14 & 16 & 18 & 20 \\
\hline
$q$ & 0.932 & 1.00 & 0.413 & 0.538 & 0.522 & 0.378 & 0.330 & 0.145 & 0.111 \\ 
$N_0$ & 6 & 2 & 6 & 6 & 34 & 2 & 2 & 2 & 2\\
$N_1$ & 8 & 12 & 14 & 14 & 126 & 48 & 48 & 172 & 158\\
$\frac{N_0}{N_0+N_1}$ & 0.429 & 0.143 & 0.300 & 0.300 & 0.213 & 0.040 & 0.040 & 0.011 & 0.013 \\
\hline\hline
\end{tabular}
\label{tbl:worst_case}
\end{table}

\begin{figure}[t]
\centering
\begin{tabular}{cc}
\epsfig{figure = 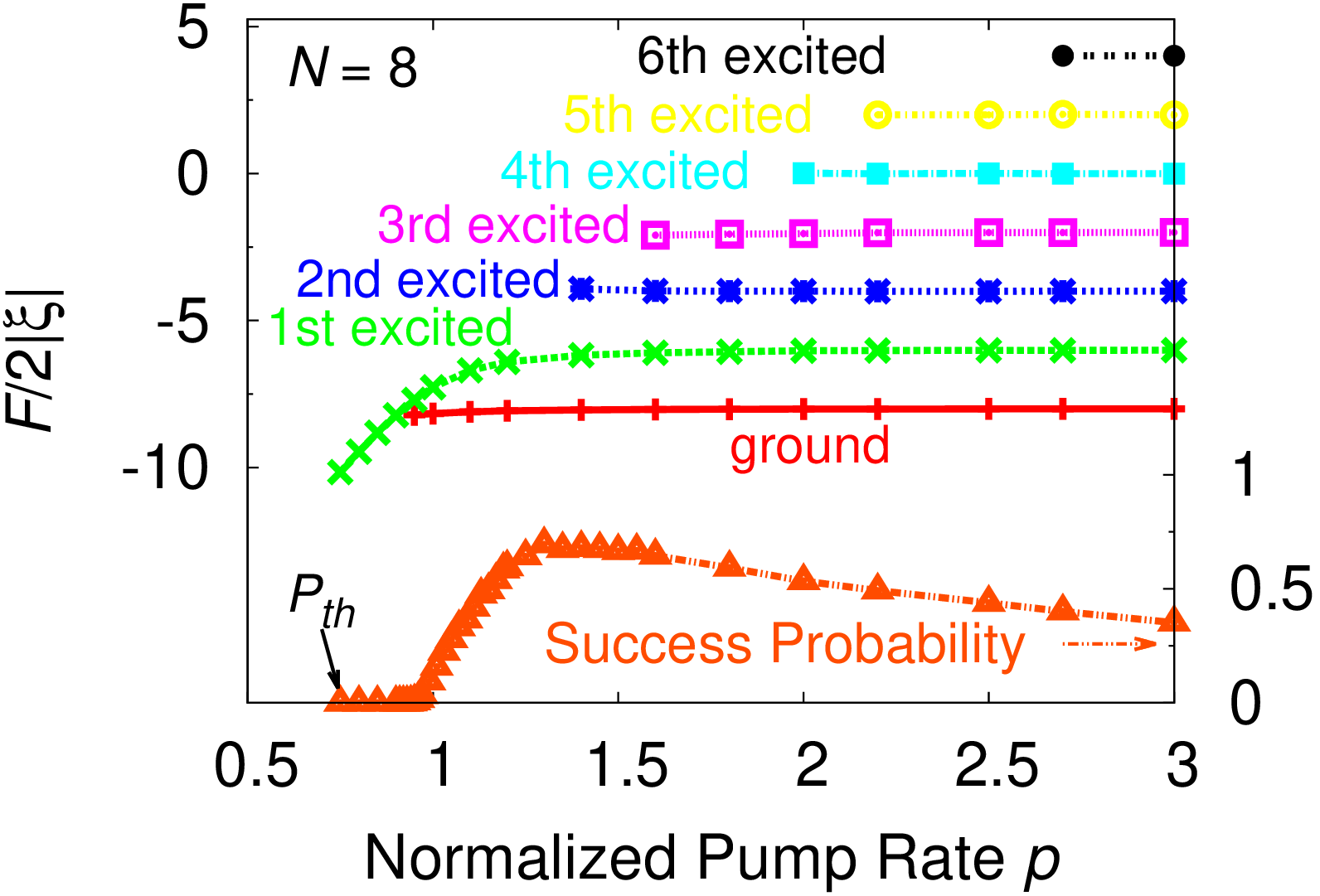, width = 1.8in, angle = -90} \hspace{0.2in} &  \hspace{0.2in} \epsfig{figure = 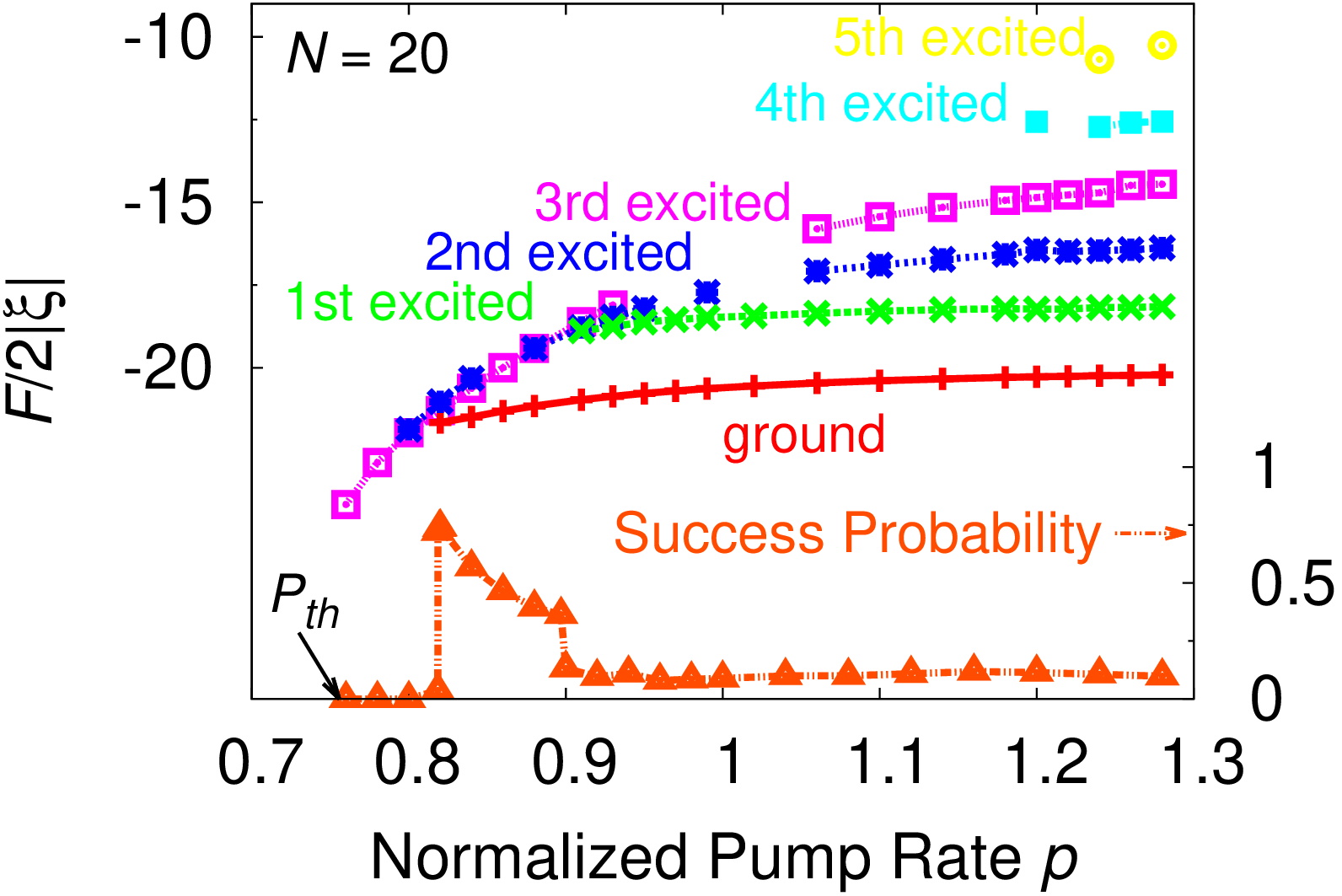, width = 1.8in, angle = -90}
\end{tabular}
\caption{The dependence of the success probability and the reachable steady states of the network on the normalized pump rate when $\xi = -0.1$ for the worst-case instances with 8 and 20 vertices.}
\label{fig:suc_pr_vs_p}
\end{figure} 

The results displayed in Fig.\ref{fig:suc_pr_vs_node_p_1.1_xi_0.1} are by no means the optimum performance of the network. Another proper combination of $p$ and $\xi$ can boost the success probability significantly. Fig.\ref{fig:suc_pr_vs_p} demonstrates the improvement in the success probability for two of the worst-case instances listed in Table \ref{tbl:worst_case}. These worst-case instances are considered to be hard ones in the MAX-CUT problem because the ratios between the number of maximum cuts and the number of second largest cuts are relatively small, especially when the graph order is large. Nevertheless, for each of the two hard instances shown in Fig.\ref{fig:suc_pr_vs_p}, there exists some optimal pump rate $p^{*}$ given a fixed coupling coefficient at which the success rate is maximized and is raised up to above 0.7.

The variation of the success probability with the normalized pump rate is closely related to the reachable steady states of the network. This relation is demonstrated in Fig.\ref{fig:suc_pr_vs_p} as well. The approached steady states of the network are classified according to the weight of the cuts they represent. Since the Ising problem and the MAX-CUT problem are mutually reducible, the maximum cut classification is labeled as ``ground", and the second largest cut classification is labeled as ``first excited'' and so on. Each classification is then associated with a function $F = \Gamma-N$. Physically, $F$ means the increased amount of the overall photon decay rate of a global mode due to the mutual coupling. If the approximation in eq.(\ref{eq:tot_gamma_Ising}) holds, $F/2|\xi|$ is exactly the Ising Hamiltonian corresponding to the MAX-CUT problem. For the two instances shown here, when the network is pumped just above its threshold, the first appearing steady states represent incorrect solutions. The success probability therefore vanishes because of the large discrepancy between $F/2|\xi|$ and the energy of the corresponding Ising spin states. The normalized pump rate $p$ has to rise to a certain level before steady states corresponding to the correct solutions can be developed. The success probability increases until $p$ reaches $p^{*}$ but drops again when the network is further pumped. This is because more and more newly reachable steady states are mapped to the incorrect solutions.

\section{Summary and Discussion}\label{sec:6}

The potential for solving the NP-hard Ising problem using a degenerate OPO network has been investigated. When the network is pumped above its threshold as defined in eq.(\ref{eq:thres_cond}) and the condition in eq.(\ref{eq:quad_0}) is satisfied, it becomes possible to convert each phase configuration of the network to an Ising state. If the mutual coupling is sufficiently weak, the overall photon decay rate of the network is proven to be in proportion to the energy of the corresponding Ising spin state as shown in eq.(\ref{eq:tot_gamma_Ising}). Even though the presented argument cannot be generalized to an arbitrary relation between the normalized pump rate and the coupling strength, eq.(\ref{eq:tot_gamma_Ising}) is considered to be valid under a less strict presumption. The evidence is given by the values of $F/2|\xi|$ shown in Fig.\ref{fig:suc_pr_vs_p}. It is not difficult to distinguish regions that contradict with the assumption for deriving eq.(\ref{eq:tot_gamma_Ising}), but $F/2|\xi|$ still gives a good approximation to the corresponding Ising energy.      

Performance of the network has been numerically studied through conducting computational experiments using the equivalent MAX-CUT problem on all cubic graphs of order up to 20. With both the normalized pump rate and the coupling coefficient being fixed at $p = 1.1$ and $\xi = -0.1$, the network has achieved a high success rate in finding a maximum cut on average. Even though the success probability for the worst-case instances decreases with the number of vertices, it can be significantly improved by a proper choice of $p$ and $\xi$ dependent on the given instance. If an efficient algorithm is discovered in finding the optimal combination, an upgrade of the worst-case performance of the network can be greatly expected.

In addition, it is worthwhile mentioning that the Ising machine based on a laser network \cite{UT11,TU12} fails in solving the worst-case instances listed in Table \ref{tbl:worst_case} when the order is above 6 without a self-learning algorithm \cite{W12}. The remarkable performance superiority of the degenerate OPO network over the laser network stems from the presence of the binary regeneration mechanism. In contrast to the bistability of the phase of a degenerate OPO, a laser diffuses its phase continuously over the interval $[0,2\pi]$.

For the convenience of theoretical description, the proposed network is constructed from spatially separated degenerate OPOs. In reality, technical issues in connecting the degenerate OPOs can be resolved by a time division multiplexing scheme \cite{MY13}. In this scheme, all the degenerate OPOs share the same optical cavity with the signal fields being amplified at different time slots, and mutual connections are implemented by coherently feeding back the output signals through optical delay lines. A network of 4 degenerate OPOs has been implemented using this technique and is able to solve the MAX-CUT problem on the cubic graph with 4 vertices successfully, which will be reported in another venue.

Performance of the network in solving large-size instances will be one of the future topics of the numerical study. Preliminary investigation has demonstrated promising results: the network can easily find solutions to benchmark instances of the MAX-CUT problem \cite{HR00} better than the 0.878-performance guarantee of the celebrated approximation algorithm based on semidefinite programming \cite{GW95}. It will be interesting to compare the outputs from the network with currently best-known solutions generated from other highly optimized methods.    

\section*{Acknowledgments}
Zhe Wang is grateful for the support from Stanford Graduate Fellowship. This project is supported by the FIRST program of Japanese government.

\bibliographystyle{ieeetr}
\bibliography{References}

\end{document}